\documentclass[aps,showpacs,preprintnumbers,pacs,nofootinbib,preprint]{revtex4}
\usepackage{graphicx}

\begin{document}

\title{The virial theorem and the dynamics of clusters of galaxies in the brane
world models}
\author{T. Harko}
\email{harko@hkucc.hku.hk}\affiliation{Department of Physics and
Center for Theoretical and Computational Physics, The University
of Hong Kong, Pok Fu Lam Road, Hong Kong SAR, P. R. China}

\author{K. S. Cheng}
\email{hrspksc@hkucc.hku.hk}\affiliation{Department of Physics and
Center for Theoretical and Computational Physics, The University
of Hong Kong, Pok Fu Lam Road, Hong Kong SAR, P. R. China}

\begin{abstract}
A version of the virial theorem, which takes into account the
effects of the non-compact extra-dimensions, is derived in the
framework of the brane world models. In the braneworld scenario,
the four dimensional effective Einstein equation has some extra
terms, called dark radiation and dark pressure, respectively,
which arise from the embedding of the 3-brane in the bulk. To
derive the generalized virial theorem we use a method based on the
collisionless Boltzmann equation. The dark radiation term
generates an equivalent mass term (the dark mass), which gives an
effective contribution to the gravitational energy. This term may
account for the well-known virial theorem mass discrepancy in
actual clusters of galaxies. An approximate solution of the vacuum
field equations on the brane, corresponding to weak gravitational
fields, is also obtained, and the expressions for the dark
radiation and dark mass are derived. The qualitative behavior of
the dark mass is similar to that of the observed virial mass in
clusters of galaxies. We compare our model with the observational
data for galaxy clusters, and we express all the physical
parameters of the model in terms of observable quantities. In
particular, we predict that the dark mass must extend far beyond
the presently considered virial radius. The behavior of the galaxy
cluster velocity dispersion in brane world models is also
considered. Therefore the study of the matter distribution and
velocity dispersion at the extragalactic scales could provide an
efficient method for testing the multi-dimensional physical
models.
\end{abstract}


\date{\today}

\maketitle

\section{Introduction}

Several recent astrophysical observations \cite{Ri98} have provided the
astonishing result that around $95-96\%$ of the content of the universe is
in the form of dark matter $+$ dark energy, with only about $4-5\%$ being
represented by baryonic matter. More intriguing, around $70\%$ of the
energy-density may be in the form of what is called the dark energy, which
provides $\Omega _{DE}\sim 0.7$, and may be responsible for the acceleration
of the distant type Ia supernovae. The best candidate for the dark energy is
the cosmological constant $\Lambda $, which is usually interpreted
physically as a vacuum energy, with energy density $\rho _{\Lambda }$ and
pressure $p_{\Lambda }$ satisfying the unusual equation of state $\rho
_{\Lambda }=-p_{\Lambda }/c^{2}=\Lambda /8\pi G$. Its size is of the order $%
\Lambda \approx 3\times 10^{-56}$ cm$^{-2}$ \cite{PeRa03}. The
analysis of the cosmic microwave radiation background data from
the WMAP experiment indicated that most of the $22\%$ dark matter
content of the Universe must be of non-baryonic nature. More
specifically, the analysis of the power spectrum indicates that a
theory of gravity based essentially on the properties of baryonic
matter would produce a lower third peak \cite{Sp06}. The same
results can also be obtained from the angular-size - redshift
relation for ultra-compact radio sources \cite{Ja1}.

The problem of the dark matter is a long standing problem in modern
astrophysics. Two important observational issues, the behavior of the
galactic rotation curves and the mass discrepancy in clusters of galaxies
led to the necessity of considering the existence of the dark matter at a
galactic and extra-galactic scale.

The rotation curves of spiral galaxies \cite{Bi87} are one of the best
evidences showing the problems Newtonian mechanics and/or standard general
relativity has to face on the galactic/intergalactic scale. In these
galaxies neutral hydrogen clouds are observed at large distances from the
center, much beyond the extent of the luminous matter. Since the clouds move
in circular orbits with velocity $v_{tg}(r)$, the orbits are maintained by
the balance between the centrifugal acceleration $v_{tg}^2/r$ and the
gravitational attraction force $GM(r)/r^2$ of the total mass $M(r)$
contained within the orbit. This allows the expression of the mass profile
of the galaxy in the form $M(r)=rv_{tg}^2/G$.

Observations show that the rotational velocities increase near the center of
the galaxy and then remain nearly constant at a value of $v_{tg\infty }\sim
200$ km/s \cite{Bi87}. This leads to a mass profile $M(r)=rv_{tg\infty }^2/G$%
. Consequently, the mass within a distance $r$ from the center of the galaxy
increases linearly with $r$, even at large distances where very little
luminous matter can be detected.

The second astrophysical evidence for dark matter comes from the study of
the clusters of galaxies. The total mass of a cluster can be estimated in
two ways. Knowing the motions of its member galaxies, the virial theorem
gives one estimate, $M_{V}$, say. The second is obtained by separately
estimating the mass of each individual members, and summing these masses, to
give a total baryonic mass $M$. Almost without exception it is found that $%
M_{V}$ is considerably greater than $M$, $M_{V}>M$, typical values of $%
M_{V}/M$ being about 20-30 \cite{Bi87}.

These behaviors at a galactic/extra-galactic scale are usually
explained by postulating the existence of some dark (invisible)
matter, distributed in a spherical halo around the galaxies. The
dark matter is assumed to be a cold, pressureless medium. There
are many possible candidates for dark matter, the most popular
ones being the weekly interacting massive particles (WIMP). Their
interaction cross section with normal baryonic matter, while
extremely small, are expected to be non-zero and we may expect to
detect them directly. It has also been suggested that the dark
matter in the Universe might be composed of superheavy particles,
with mass $\geq 10^{10}$ GeV. But observational results show the
dark matter can be composed of superheavy particles only if these
interact weakly with normal matter or if their mass is above
$10^{15}$ GeV \cite{AlBa03}.

Dark matter may also be in the form of Warm Dark Matter particles,
consisting either of sterile neutrinos, with masses of several
keV, or from early decoupled thermal relics \cite{ster}. However,
constraints based on X-ray fluxes from the Andromeda galaxy have
shown that dark matter particles cannot be sterile neutrinos,
unless they are produced by a nonstandard mechanism (resonant
oscillations, coupling with the inflaton field), or they have been
diluted by some large entropy release. The X-rays produced by the
decays of these relic sterile neutrinos can boost the production
of molecular hydrogen, which can speed up the cooling of gas and
the early star formation \cite{ster}.

Hence, despite more than 20 years of intense experimental and
observational effort, up to now no \textit{non-gravitational}
evidence for dark matter has ever been found: no direct evidence
of it and no annihilation radiation from it. Moreover, accelerator
and reactor experiments do not support the physics (beyond the
standard model) on which the dark matter hypothesis is based upon.

Therefore, it seems that the possibility that Einstein's (and the Newtonian)
gravity breaks down at the scale of galaxies cannot be excluded \textit{a
priori}. Several theoretical models, based on a modification of Newton's law
or of general relativity, have been proposed to explain the behavior of the
galactic rotation curves \cite{dark}.

The proposal by Randall and Sundrum \cite{RS99a} that our four-dimensional
space-time is a three-brane, embedded in a five-dimensional space-time (the
bulk), had attracted a considerable interest lately. According to the
brane-world scenario, the physical fields (electromagnetic, Yang-Mills etc.)
in our four-dimensional Universe are confined to the three brane. These
fields are assumed to arise as fluctuations of branes in string theories.
Only gravity can freely propagate in both the brane and bulk space-times,
with the gravitational self-couplings not significantly modified. Even if
the fifth dimension is uncompactified, standard $4D$ gravity is reproduced
on the brane. Hence this model allows the presence of large, or even
infinite non-compact extra dimensions. Our brane is identified to a domain
wall in a $5$-dimensional anti-de Sitter space-time. For a review of
dynamics and geometry of brane Universes see \cite{Ma01}.

Due to the correction terms coming from the extra dimensions, significant
deviations from the Einstein theory occur in brane world models at very high
energies \cite{SMS00}. Gravity is largely modified at the electro-weak scale
$1$ TeV. The cosmological and astrophysical implications of the brane world
theories have been extensively investigated in the physical literature \cite
{all2}.

The vacuum field equations on the brane have been reduced to a
system of two ordinary differential equations, which describe all
the geometric properties of the vacuum as functions of the dark
pressure and dark radiation terms (the projections of the Weyl
curvature of the bulk, generating non-local brane stresses) in
\cite{Ha03}. Several classes of vacuum solutions of the static
gravitational field equations in the brane world scenario
describing conformally symmetric gravitational fields in the
vacuum have been obtained in \cite{Ma04,Ha04}. As a possible
physical application of these solutions the behavior of the
angular velocity $v_{tg}$ of the test particles in stable circular
orbits on the brane has been considered. In the obtained models in
the limit of large radial distances and for a particular set of
values of the integration constants the angular velocity tends to
a constant value. This behavior is typical for massive particles
(hydrogen clouds) outside galaxies \cite{Bi87}, and is usually
explained by postulating the existence of the dark matter. Thus,
the rotational galactic curves can be naturally explained in brane
world models without introducing any additional hypothesis. The
galaxy is embedded in a modified, spherically symmetric geometry,
generated by the non-zero contribution of the Weyl tensor from the
bulk. The extra-terms, which can be described in terms of the dark
radiation term $U$ and the dark pressure term $P$, act as a
"matter" distribution outside the galaxy
\cite{Ma04,Ha04,Ha05,BoHa07}.

The existence of the dark radiation term generates an equivalent mass term $%
M_{U}$, which is linearly increasing with the distance, and proportional to
the baryonic mass of the galaxy $M_{B}$, $M_{U}(r)\approx M_{B}(r/r_{0})$.
The particles moving in this geometry feel the gravitational effects of $U$,
which can be expressed in terms of an equivalent mass. In the framework of
this model all the relevant physical parameters (metric tensor components,
dark radiation and dark pressure terms) can be obtained as functions of the
tangential velocity, and hence they can be determined observationally.
Using the smallness of the rotational velocity, a perturbative scheme for reconstructing the metric in a galactic halo in
the braneworld models with induced gravity
was developed in \cite{VySh07}. In the leading order of expansion, at sufficiently large distances, the obtained result reproduce that obtained in the Randall-Sundrum braneworld model \cite{Ma04,Ha04,Ha05,BoHa07}.

Similar interpretations of the dark matter as bulk effects have been also
considered in \cite{Pal}.

The exact galactic metric, the dark radiation and the dark pressure in the
flat rotation curves region in the brane world scenario has been obtained in
\cite{Ha05}. The deflection of photons has also been considered and the
bending angle of light is computed. The bending angle predicted by the brane
world models is much larger than that predicted by standard general
relativistic and dark matter models. Therefore the study of the light
deflection by galaxies and the gravitational lensing could discriminate
between the different dynamical laws proposed to model the motion of
particles at the galactic level and the standard dark matter models.

Because of its generality and wide range of applications, the virial theorem
plays an important role in astrophysics. Assuming steady state, one of the
important results which can be obtained with the use of the virial theorem
is to deduce the mean density of astrophysical objects such as galaxies,
clusters and super clusters, by observing the velocities of test particles
rotating around them. Hence the virial theorem can be used to predict the
total mass of the clusters of galaxies. The virial theorem is also a
powerful tool for stability studies. In a general relativistic framework
several versions of the virial theorem have been proposed \cite{vir}.
Versions of the virial theorem including the effect of a cosmological
constat have been derived in \cite{Ja70} and \cite{No}, respectively.

It is the purpose of the present paper to consider the virial
theorem in the framework of the brane world models. By taking into
account the effects of the non-compact extra-dimensions we derive,
by using the collisionless Boltzmann equation, a generalized
virial equality, which explicitly takes into account the presence
of the bulk effects. In the case of static spherically symmetric
systems the transmitted projections of the bulk Weyl tensor can be
described in terms of two quantities, called the dark radiation
and the dark pressure, respectively. The dark radiation term gives
an effective contribution to the gravitational energy, the total
virial mass being proportional to the mass associated to the
effective mass of the dark radiation. This term may account for
the well-known virial theorem mass discrepancy in actual clusters
of galaxies. The generalized virial theorem in the brane world
models can be an efficient tool in observationally testing the
brane world models.

An approximate solution of the vacuum field equations on the
brane, corresponding to weak gravitational fields, is also
obtained, and the expressions for the dark radiation and dark mass
are derived. The qualitative behavior of the dark mass is similar
to that of the observed virial mass in clusters of galaxies.

We consider several astrophysical applications of the general
results obtained in the present paper. The expressions of the dark
radiation and of the dark mass can be expressed in terms of
observable astrophysical quantities, like the temperature and the
gas density profile, which can be obtained from the X-ray
observations of the gas in the cluster. From the obtained
expression of the dark radius, which describes the relevant length
scale for brane world effects, we predict that the dark mass must
extend far beyond the presently considered virial radius. The
behavior of the galaxy cluster velocity dispersion in brane world
models is also considered in detail. Therefore the study of the
matter distribution and velocity dispersion at the extragalactic
scales could provide an efficient method for testing the
multi-dimensional physical models.

The present paper is organized as follows. The basic equations for
static spherically symmetric gravitational fields on the brane are
presented in Section II. The generalized virial theorem, including
the effects of the non-compact extra dimensions is derived in
Section III. In Section IV we derive the general expressions of
the cluster metric, of the dark radiation and of the dark mass for
the clusters of galaxies. Some astrophysical applications are
presented in Section V. We discuss and conclude our results in
Section VI.

\section{Static spherically symmetric gravitational fields on the brane}

On the $5$-dimensional space-time (the bulk), with the negative vacuum
energy $\Lambda _{5}$ and brane energy-momentum as source of the
gravitational field, the Einstein field equations are given by
\begin{equation}
G_{IJ}=k_{5}^{2}T_{IJ},\qquad T_{IJ}=-\Lambda _{5}g_{IJ}+\delta (Y)\left[
-\lambda _{b}g_{IJ}+T_{IJ}^{\text{matter}}\right] ,
\end{equation}
with $\lambda _{b}$ the vacuum energy on the brane and $k_{5}^{2}=8\pi G_{5}$%
. In this space-time a brane is a fixed point of the $Z_{2}$ symmetry. In
the following capital Latin indices run in the range $0,...,4$, while Greek
indices take the values $0,...,3$.

Assuming a metric of the form
$ds^{2}=(n_{I}n_{J}+g_{IJ})dx^{I}dx^{J}$, with $n_{I}dx^{I}=d\chi
$ the unit normal to the $\chi =$constant hypersurfaces and
$g_{IJ}$ the induced metric on $\chi =$constant hypersurfaces, the
effective four-dimensional gravitational equations on the brane
 take the form \cite{SMS00}:
\begin{equation}
G_{\mu \nu }=-\Lambda g_{\mu \nu }+k_{4}^{2}T_{\mu \nu }+k_{5}^{4}S_{\mu \nu
}-E_{\mu \nu },  \label{Ein}
\end{equation}
where $S_{\mu \nu }$ is the local quadratic energy-momentum correction
\begin{equation}
S_{\mu \nu }=\frac{1}{12}TT_{\mu \nu }-\frac{1}{4}T_{\mu }{}^{\alpha }T_{\nu
\alpha }+\frac{1}{24}g_{\mu \nu }\left( 3T^{\alpha \beta }T_{\alpha \beta
}-T^{2}\right) ,
\end{equation}
and $E_{\mu \nu }$ is the non-local effect from the free bulk gravitational
field, the transmitted projection of the bulk Weyl tensor $C_{IAJB}$, $%
E_{IJ}=C_{IAJB}n^{A}n^{B}$, with the property $E_{IJ}\rightarrow
E_{\mu \nu }\delta _{I}^{\mu }\delta _{J}^{\nu }\quad $as$\quad
\chi \rightarrow 0$. We have also denoted $k_{4}^{2}=8\pi G$, with
$G$ the usual four-dimensional gravitational constant. Eq.
(\ref{Ein}) is also known as the effective Einstein equation.  It
has been shown that for a large class of generalized
Randall-Sundrum type models, the characterization of brane-gravity
sector by the effective Einstein equation, Codazzi equation and
the twice-contracted Gauss equation is equivalent with the bulk
five-dimensional Einstein equation \cite{Ge03}.

The four-dimensional cosmological constant, $\Lambda $, and the
four-dimensional coupling constant, $k_{4}$, are given by $\Lambda
=k_{5}^{2}\left( \Lambda _{5}+k_{5}^{2}\lambda _{b}^{2}/6\right) /2$ and $%
k_{4}^{2}=k_{5}^{4}\lambda _{b}/6$, respectively. The
four-dimensional gravitational constant $G$ is given by
$G=k_5^4\lambda _b/48\pi $ \cite{SMS00}.

The Einstein equation in the bulk and the Codazzi equation also imply the
conservation of the energy-momentum tensor of the matter on the brane, $%
D_{\nu }T_{\mu }{}^{\nu }=0$, where $D_{\nu }$ denotes the brane covariant
derivative. Moreover, from the contracted Bianchi identities on the brane it
follows that the projected Weyl tensor should obey the constraint $D_{\nu
}E_{\mu }{}^{\nu }=k_{5}^{4}D_{\nu }S_{\mu }{}^{\nu }$ \cite{Ma01}.

In the limit $\lambda _{b}^{-1}\rightarrow 0$ we recover standard
general relativity. An alternative possibility in recovering the
four-dimensional Einstein equations is to take the limit
$k_5\rightarrow 0$, while keeping the Newtonian gravitational
constant $G$ finite \cite{SMS00}. As one can see from its
definition, the existence of the four-dimensional gravitational
constant relies on the existence of the vacuum energy $\lambda
_b$. It is impossible to define an the gravitational constant in
an era when the distinction between vacuum energy and matter is
ambiguous. Moreover, the positivity of $G$ also requires the
positivity of $\lambda _b$ \cite{SMS00}.

The symmetry properties of $E_{\mu \nu }$ imply that in general we
can decompose it irreducibly with respect to a chosen $4$-velocity
field $u^{\mu }$ as
\begin{equation}
E_{\mu \nu }=-\tilde{k}^{4}\left[ U\left( u_{\mu }u_{\nu }+\frac{1}{3}h_{\mu
\nu }\right) +2Q_{(\mu }u_{\nu )}+P_{\mu \nu }\right] ,  \label{WT}
\end{equation}
where $\tilde{k}=k_{5}/k_{4}$, $h_{\mu \nu }=g_{\mu \nu }+u_{\mu }u_{\nu }$
projects orthogonal to $u^{\mu }$, the ''dark radiation'' term $U=-\tilde{k}%
^{-4}E_{\mu \nu }u^{\mu }u^{\nu }$ is a scalar, $Q_{\mu
}=\tilde{k}^{-4}h_{\mu
}^{\alpha }E_{\alpha \beta }u^{\beta }$ a spatial vector and $P_{\mu \nu }=-\tilde{k}%
^{-4}\left[ h_{(\mu }\text{ }^{\alpha }h_{\nu )}\text{ }^{\beta }-\frac{1}{3}%
h_{\mu \nu }h^{\alpha \beta }\right] E_{\alpha \beta }$ a spatial,
symmetric and trace-free tensor \cite{Ma01}.

We assume that the matter on the brane consists of an anisotropic fluid,
characterized by an effective energy-density $\rho _{eff}\neq 0$, a radial
pressure $p_{eff}^{(r)}$ and a tangential pressure $p_{eff}^{\left( \perp
\right) }$, respectively. Generally $p_{eff}^{(r)}\neq $ $p_{eff}^{\left(
\perp \right) }$, but for isotropic systems $p_{eff}^{(r)}=$ $%
p_{eff}^{\left( \perp \right) }$. Hence $T_{\mu \nu }\neq 0$, and
consequently also $S_{\mu \nu }\neq 0$. $E_{\mu \nu }$ satisfies the
constraint $D_{\nu }E_{\mu }{}^{\nu }=k_{5}^{4}D^{\mu }S_{\mu \nu }$, where $%
D_{\mu }$ is the projection (orthogonal to $u^{\mu }$) of the covariant
derivative \cite{SMS00}. In an inertial frame at any point on the brane we
have $u^{\mu }=\delta _{0}^{\mu }$ and $h_{\mu \nu }=$diag$\left(
0,1,1,1\right) $. In the static spherically symmetric case $Q_{\mu }=0$ and
the constraint for $E_{\mu \nu }$ takes the form \cite{GeMa01}
\begin{equation}
\frac{1}{3}D_{\mu }U+\frac{4}{3}UA_{\mu }+D^{\nu }P_{\mu \nu }+A^{\nu
}P_{\mu \nu }=-\left( 4\pi G\right) ^{2}\left( 2\rho
_{eff}+p_{eff}^{(r)}+p_{eff}^{\left( \perp \right) }\right) D_{\mu }\rho
_{eff},
\end{equation}
where $A_{\mu }=u^{\nu }D_{\nu }u_{\mu }$ is the 4-acceleration.

In the static spherically symmetric case we may choose $A_{\mu }=A(r)r_{\mu
} $ and $P_{\mu \nu }=P(r)\left( r_{\mu }r_{\nu }-\frac{1}{3}h_{\mu \nu
}\right) $, where $A(r)$ and $P(r)$ (the ''dark pressure'') are some scalar
functions of the radial distance $r$, and $r_{\mu }$ is a unit radial vector
\cite{Da00}.

We choose the static spherically symmetric metric on the brane in the form
\begin{equation}
ds^{2}=-e^{\nu \left( r\right) }dt^{2}+e^{\lambda \left( r\right)
}dr^{2}+r^{2}\left( d\theta ^{2}+\sin ^{2}\theta d\phi ^{2}\right) .
\label{line}
\end{equation}

Then the gravitational field equations and the effective energy-momentum
tensor conservation equation for an anisotropic static spherically symmetric
system take the form \cite{Ha03,Ha04,Ma04,Ha05,GeMa01}
\begin{equation}
-e^{-\lambda }\left( \frac{1}{r^{2}}-\frac{\lambda ^{\prime }}{r}\right) +%
\frac{1}{r^{2}}=8\pi G\rho _{eff}\left(1+\frac{\rho _{eff}}{2\lambda _b}%
\right)+\frac{48\pi G}{k_4^{4}\lambda _{b}}U+\Lambda ,  \label{f1}
\end{equation}
\begin{equation}
e^{-\lambda }\left( \frac{\nu ^{\prime }}{r}+\frac{1}{r^{2}}\right) -\frac{1%
}{r^{2}}=8\pi Gp_{eff}^{(r)}+\frac{4\pi G}{\lambda _b}\rho _{eff}\left(\rho
_{eff}+2p_{eff}^{(r)}\right)+\frac{16\pi G}{k_4^{4}\lambda _{b}}\left(
U+2P\right)-\Lambda ,  \label{f2}
\end{equation}
\begin{equation}
e^{-\lambda }\left( \nu ^{\prime \prime }+\frac{\nu ^{\prime 2}}{2}+\frac{%
\nu ^{\prime }-\lambda ^{\prime }}{r}-\frac{\nu ^{\prime }\lambda ^{\prime }%
}{2}\right) =16\pi Gp_{eff}^{\left(\perp \right))}+\frac{8\pi G}{\lambda _b}%
\rho _{eff}\left(\rho _{eff}+2p_{eff}^{\left(\perp \right)}\right)+\frac{%
32\pi G}{k_4^{4}\lambda _{b}}\left( U-P\right)-2\Lambda ,  \label{f3}
\end{equation}
\begin{equation}
\nu ^{\prime }=-\frac{U^{\prime }+2P^{\prime }}{2U+P}-\frac{6P}{r\left(
2U+P\right) }-\left(4\pi G\right)^2\frac{2\rho
_{eff}+p_{eff}^{(r)}+p_{eff}^{\left(\perp \right)}}{2U+P}\rho _{eff}^{\prime
}.  \label{f4}
\end{equation}

In the following we shall denote $\alpha =16\pi G/k_4^{4}\lambda _{b}=1/4\pi
G\lambda _b$.

\section{The virial theorem in the brane world models}

We consider an isolated, spherically symmetric cluster, situated in a space
with metric given by Eq. (\ref{line}). We describe the galaxies, which are
treated as identical, collisionless point particles, by a distribution
function which obeys the general relativistic Boltzmann equation.

Consider a time-oriented Lorentzian four-dimensional space-time manifold $M$%
, with metric $g$ of signature $\left( -,+,+,+\right) $. The tangent bundle $%
T(M)$ is a real vector bundle whose fibers at a point $x\in M$ is given by
the tangent space $T_{x}\left( M\right) $. In the space-time $M$ the
instantaneous state of a particle with mass $m_0$ is given by a
four-momentum $p\in T_{x}\left( M\right) $ at an event $x\in M$. The
one-particle phase space $P_{phase}$ is a subset of the tangent bundle given
by \cite{Li66}
\begin{equation}
P_{phase}:=\left\{ \left( x,p\right)\left|\right. x\in M,p\in
T_{x}\left( M\right) ,p^{2}=-m_0^2\right\} .
\end{equation}

A state of a multi-particle system is described by a continuous,
non-negative distribution function $f\left( x,p\right) $, defined on $%
P_{phase}$, and which gives the number $dN$ of the particles of the system
which cross a certain space-like volume $dV$ at $x$, and whose 4-momenta $p$
lie within a corresponding three-surface element $d\vec{p}$ in the momentum
space. The mean value of $f$ gives the average number of occupied particle
states $\left( x,p\right) $. Macroscopic, observable quantities can be
defined as moments of $f$ \cite{Li66}.

Let $\left\{ x^{\alpha }\right\} $, $\alpha =0,1,2,3$ be a local
coordinate system in $M$, defined in some open set $U\subset M$.
The coordinates are chosen so that $\partial _t$ is time-like
future directed and $\partial _a$, $a=1,2,3$, are spacelike. Then
$\left\{\partial /\partial x^{\alpha }\right\} $ is the
corresponding natural basis for tangent vectors. We express each
tangent vector $p$ in $U$ in terms of this basis as $p=p^{\alpha
}\partial /\partial x^{\alpha }$ and define a
system of local coordinates $\left\{ z^{A}\right\} $, $A=0,...,7$ in $%
T_{U}(M)$ as $z^{\alpha }=x^{\alpha }$, $z^{\alpha +4}=p^{\alpha }$. This
defines a natural basis in the tangent space given by $\left\{ \partial
/\partial z^{A}\right\} =\left\{ \partial /\partial x^{\alpha },\partial
/\partial p^{\alpha }\right\} $ \cite{Li66}.

A vertical vector field over $TM$ is given by $\pi =p^{\alpha }\partial
/\partial p^{\alpha }$. The geodesic flow field $\sigma $, which can be
constructed over the tangent bundle, is defined as $\sigma =p^{\alpha
}\partial /\partial x^{\alpha }-p^{\alpha }p^{\gamma }\Gamma _{\alpha \gamma
}^{\beta }\partial /\partial p^{\beta }=p^{\alpha }D_{\alpha }$, where $%
D_{\alpha }=\partial /\partial x^{\alpha }-p^{\gamma }\Gamma
_{\alpha \gamma }^{\beta }\partial /\partial p^{\beta }$, and
$\Gamma _{\alpha \gamma }^{\beta }$ are the connection
coefficients. Physically, $\sigma $ describes the phase flow for a
stream of particles whose motion through space-time is geodesic
\cite{Li66}.

Therefore the transport equation for the propagation of a particle in a
curved arbitrary Riemannian space-time is given by the Boltzmann equation
\cite{Li66}
\begin{equation}  \label{distr}
\left( p^{\alpha }\frac{\partial }{\partial x^{\alpha }}-p^{\alpha }p^{\beta
}\Gamma _{\alpha \beta }^{i}\frac{\partial }{\partial p^{i}}\right) f=0.
\end{equation}

For many applications it is convenient to introduce an appropriate
orthonormal frame or tetrad $e_{\mu }^{(a)}\left( x\right) $,
$a=0,1,2,3$, which varies smoothly over some coordinates
neighborhood $U$ and satisfies the condition $e_{\mu }^{(a)}\left(
x\right) e_{\mu }^{(b)}\left( x\right) =\eta ^{ab}$ for all $x\in
U$ \cite{Li66,Ja70}. Any tangent vector $p^{\mu }$ at $x$ can be
expressed as $p^{\mu }=p^{a}e_{(a)}^{\mu }$, which defines the
tetrad components $p^{a}$.

In the case of the spherically symmetric line element given by Eq. (\ref
{line}) we introduce the following frame of orthonormal vectors \cite
{Li66,Ja70}:
\begin{equation}
e_{\mu }^{(0)}=e^{\nu /2}\delta _{\mu }^{0},e_{\mu }^{(1)}=e^{\lambda
/2}\delta _{\mu }^{1},e_{\mu }^{(2)}=r\delta _{\mu }^{2},e_{\mu
}^{(3)}=r\sin \theta \delta _{\mu }^{3}.
\end{equation}

Let $u^{\mu }$ be the four-velocity of a typical galaxy,
satisfying the condition $u^{\mu }u_{\mu }=-1$, with tetrad
components $u^{(a)}=u^{\mu }e_{\mu }^{(a)}$. The relativistic
Boltzmann equation in tetrad components is
\begin{equation}
u^{(a)}e_{(a)}^{\mu }\frac{\partial f}{\partial x^{\mu }}+\gamma
_{(b)(c)}^{(a)}u^{(b)}u^{(c)}\frac{\partial f}{\partial u^{(a)}}=0,
\label{tetr}
\end{equation}
where the distribution function $f=f\left( x^{\mu },u^{(a)}\right) $ and $%
\gamma _{(b)(c)}^{(a)}=e_{\mu ;\nu }^{(a)}e_{(b)}^{\mu }e_{(c)}^{\nu }$ are
the Ricci rotation coefficients \cite{Li66,Ja70}. By denoting
\begin{equation}
u^{(0)}=u_{t},u^{(1)}=u_{r},u^{(2)}=u_{\theta },u^{(3)}=u_{\phi },
\end{equation}
and by assuming that the only coordinate dependence of the distribution
function is upon the radial coordinate $r$, Eq. (\ref{tetr}) becomes \cite
{Ja70}
\begin{eqnarray}  \label{tetr1}
u_{r}\frac{\partial f}{\partial r}-\left( \frac{1}{2}u_{t}^{2}\frac{\partial
\nu }{\partial r}-\frac{u_{\theta }^{2}+u_{\phi }^{2}}{r}\right) \frac{%
\partial f}{\partial u_{r}}-\frac{1}{r}u_{r}\left( u_{\theta }\frac{\partial
f}{\partial u_{\theta }}+u_{\phi }\frac{\partial f}{\partial u_{\phi }}%
\right) -  \nonumber \\
\frac{1}{r}e^{\lambda /2}u_{\phi }\cot \theta \left( u_{\theta }\frac{%
\partial f}{\partial u_{\phi }}-u_{\phi }\frac{\partial f}{\partial
u_{\theta }}\right) =0.
\end{eqnarray}

The spherical symmetry requires that the coefficient of $\cot \theta $ be
zero, which implies that $f$ is a function of $r$, $u_{r}$ and $u_{\theta
}^{2}+u_{\phi }^{2}$ only. We multiply now Eq. (\ref{tetr1}) by $mu_{r}du $,
where $m$ is the mass of each galaxy and $du =du_{r}du_{\theta }du_{\phi
}/u_{t}$ is the invariant volume element of the velocity space. By
integrating over the velocity space and by assuming that $f$ vanishes
sufficiently rapidly as the velocities tend to $\pm \infty $, we obtain
\begin{equation}
r\frac{\partial }{\partial r}\left[ \rho \left\langle u_{r}^{2}\right\rangle %
\right] +\frac{1}{2}\rho \left[ \left\langle u_{t}^{2}\right\rangle
+\left\langle u_{r}^{2}\right\rangle \right] r\frac{\partial \nu }{\partial r%
}-\rho \left[ \left\langle u_{\theta }^{2}\right\rangle +\left\langle
u_{\phi }^{2}\right\rangle -2\left\langle u_{r}^{2}\right\rangle \right] =0,
\label{tetr2}
\end{equation}
where, at each point, $\left\langle u_{r}^{2}\right\rangle $ is the average
value of $u_{r}^{2}$ etc. and $\rho $ is the mass density.

By multiplying Eq. (\ref{tetr2}) by $4\pi r^{2}$ and integrating over the
cluster we obtain \cite{Ja70}
\begin{equation}  \label{kin}
-\int_{0}^{R}4\pi \rho \left[ \left\langle u_{r}^{2}\right\rangle
+\left\langle u_{\theta }^{2}\right\rangle +\left\langle u_{\phi
}^{2}\right\rangle \right] r^{2}dr+\frac{1}{2}\int_{0}^{R}4\pi r^{3}\rho %
\left[ \left\langle u_{t}^{2}\right\rangle +\left\langle
u_{r}^{2}\right\rangle \right] \frac{\partial \nu }{\partial r}dr=0.
\end{equation}

In terms of the distribution function the energy-momentum tensor of the
matter can be written as \cite{Li66}
\begin{equation}
T_{\mu \nu }=\int fmu_{\mu }u_{\nu }du,
\end{equation}
which gives
\begin{equation}
\rho _{eff}=\rho \left\langle u_{t}^{2}\right\rangle,p_{eff}^{(r)}=\rho
\left\langle u_{r}^{2}\right\rangle,p_{eff}^{\left(\perp \right)}=\rho
\left\langle u_{\theta }^{2}\right\rangle=\rho \left\langle u_{\phi
}^{2}\right\rangle.
\end{equation}

Then, with the use of this form of the energy-momentum tensor, by
adding the gravitational field equations Eqs.
(\ref{f1})-(\ref{f3}), we immediately obtain the following
relation:
\begin{eqnarray}  \label{ff}
e^{-\lambda }\left( \frac{\nu ^{\prime \prime }}{2}+\frac{\nu ^{\prime 2}}{4}%
+\frac{\nu ^{\prime }}{r}-\frac{\nu ^{\prime }\lambda ^{\prime }}{4}\right)
&=&3\alpha U-\Lambda +4\pi G\rho \left[ \left\langle u_{t}^{2}\right\rangle
+\left\langle u_{r}^{2}\right\rangle +\left\langle u_{\theta
}^{2}\right\rangle +\left\langle u_{\phi }^{2}\right\rangle \right] +
\nonumber \\
&&\frac{4\pi G}{\lambda _{b}}\rho ^{2}\left[ \left\langle
u_{t}^{2}\right\rangle ^{2}+\left\langle u_{r}^{2}\right\rangle
^{2}+\left\langle u_{\theta }^{2}\right\rangle ^{2}+\left\langle u_{\phi
}^{2}\right\rangle ^{2}\right] .
\end{eqnarray}

It is convenient to introduce some approximations at this moment. First of
all, we assume that $\nu $ and $\lambda $ are small, so that in Eq. (\ref{ff}%
) the quadratic terms can be neglected. Secondly, we assume that
the galaxies have velocities much smaller than the velocity of the
light, so that $\left\langle u_{r}^{2}\right\rangle ,\left\langle
u_{\theta }^{2}\right\rangle ,\left\langle u_{\phi
}^{2}\right\rangle <<\left\langle u_{t}^{2}\right\rangle \approx
1$. Since for clusters of galaxies the ratio of the matter density
and of the brane tension is much smaller than $1$, $\rho /\lambda
_b<<1$, we can neglect in Eq. (\ref{ff}) the quadratic term in the
matter density. These conditions certainly apply to test particles
in stable circular motion around galaxies, and to the galactic
clusters.

Therefore Eqs. (\ref{kin}) and (\ref{ff}) become
\begin{equation}
-2K+\frac{1}{2}\int_{0}^{R}4\pi r^{3}\rho \frac{\partial \nu }{\partial r}%
dr=0,  \label{cond1}
\end{equation}
\begin{equation}
4\pi G\rho =\frac{1}{2}\frac{1}{r^{2}}\frac{\partial }{\partial r}\left(
r^{2}\frac{\partial \nu }{\partial r}\right) +\Lambda -3\alpha U,
\label{fin1}
\end{equation}
respectively, where
\begin{equation}
K=\int_{0}^{R}2\pi \rho \left[ \left\langle u_{r}^{2}\right\rangle
+\left\langle u_{\theta }^{2}\right\rangle +\left\langle u_{\phi
}^{2}\right\rangle \right] r^{2}dr,
\end{equation}
is the total kinetic energy of the galaxies. The gravitational potential
energy $\Omega $ of the system is defined by
\begin{equation}
\Omega =-\int_{0}^{R}\frac{GM(r)}{r}dM(r),
\end{equation}
where $M(r)$ is the mass out to radius $r$, so that $dM(r)=4\pi \rho r^{2}dr$%
. The total mass of the system is given by
$M=\int_{0}^{R}dM(r)=4\pi \int_{0}^{R}\rho r^{2}dr$. The main
contribution to $M$ is represented by the baryonic mass of the
intra-cluster gas and of the stars, but other particles, like, for
example, massive neutrinos, may also give a significant
contribution to $M$.

Multiplying Eq. (\ref{fin1}) by $r^{2}$ and integrating from $0$ to $r$ we
obtain
\begin{equation}
GM(r)=\frac{1}{2}r^{2}\frac{\partial \nu }{\partial r}+\frac{1}{3}\Lambda
r^{3}-3\alpha \int_{0}^{r}U(r)r^{2}dr.  \label{fin2}
\end{equation}

In the following we denote
\begin{equation}  \label{darkmass}
GM_{U}\left( r\right) =3\alpha \int_{0}^{r}U(r)r^{2}dr.
\end{equation}

This quantity may be called as the dark mass. By multiplying Eq. (\ref{fin2}%
) with $dM(r)$, integrating from $0$ to $R$, by introducing the moment of
inertia of the system as $I=\int_{0}^{R}r^{2}dM(r)$, and by denoting
\begin{equation}
\Omega _{U}=\int_{0}^{R}\frac{GM_{U}(r)dM(r)}{r},
\end{equation}
we obtain
\begin{equation}
-\Omega =\frac{1}{2}\int_{0}^{R}4\pi r^{3}\rho \frac{\partial \nu }{\partial
r}dr+\frac{1}{3}\Lambda I-\Omega _{U}.
\end{equation}

Finally, with the use of the Eq. (\ref{cond1}), we obtain the generalization
of the virial theorem in the brane world models in the form
\begin{equation}
2K+\Omega +\frac{1}{3}\Lambda I-\Omega _{U}=0.  \label{theor}
\end{equation}

The generalized virial theorem, given by Eq. (\ref{theor}), can be written
in a simpler form if we introduce the radii $R_{V}$, $R_{I}$ and $R_{U}$
defined by
\begin{equation}
R_{V}=M^{2}/\int_{0}^{R}\frac{M(r)}{r}dM(r),
\end{equation}
\begin{equation}
R_{I}=\left[ \left( \int_{0}^{R}r^{2}dM(r)\right) /M(r)\right] ^{1/2},
\end{equation}
\begin{equation}
R_{U}=M_{U}^{2}/\int_{0}^{R}\frac{M_{U}(r)dM(r)}{r},  \label{RU3}
\end{equation}
so that
\begin{equation}
\Omega =-\frac{GM^{2}}{R_{V}},
\end{equation}
\begin{equation}
I=MR_{I}^{2},
\end{equation}
\begin{equation}
\Omega _{U}=\frac{GM_{U}^{2}}{R_{U}}.
\end{equation}

$R_U$ may be called the dark radius of the cluster of galaxies. The virial
mass $M_{V}$ is defined as \cite{Ja70}
\begin{equation}
2K=\frac{GM_{V}^{2}}{R_{V}}.
\end{equation}

If this expression is substituted into the virial theorem, given by Eq. (\ref
{theor}), we obtain
\begin{equation}
\frac{M_{V}}{M}=\left( 1+\frac{M_{U}^{2}R_{V}}{M^{2}R_{U}}-\frac{\Lambda }{%
4\pi G\bar{\rho}}\right) ^{1/2},  \label{fin6}
\end{equation}
where $\bar{\rho}=3M/4\pi R_{V}R_{I}^{2}$.

For $U=0$, $M_U=0$ and we reobtain the virial theorem in the presence of a
cosmological constant \cite{Ja70}.

If $M_{V}/M>3$, a condition which is true for most of the clusters, the term
unity can be neglected in the bracket in Eq. (\ref{fin6}) with little loss
of accuracy. Moreover, the contribution of the cosmological constant to the
mass energy of the galaxy can also be ignored as being several orders of
magnitude smaller than the observed masses. Therefore the virial mass in the
brane world models is given by
\begin{equation}
M_{V}\approx M_{U}\sqrt{\frac{R_{V}}{R_{U}}}.  \label{virial}
\end{equation}

From observational point of view the virial mass $M_V$ is
determined from the study of the velocity dispersion $\sigma _r^2$
of the stars and of the galaxies in the clusters. According to our
interpretation, most of the mass in a cluster with mass $M_{tot}$
should be in the form of the dark mass $M_U$, so that $M_U\approx
M_{tot}$. A possibility of detecting the presence of the dark mass
and of the astrophysical effects of the extra dimensions is
through gravitational lensing, which can provide direct evidence
of the mass distribution and of the gravitational effects even at
distances extending far beyond of the virial radius of the
cluster.

\section{Dark mass, dark radiation and metric for clusters of galaxies on
the brane}

The total mass of galaxy clusters ranges from $10^{13}$ $M_{\odot }$ for
groups up to a few $10^{15}$ $M_{\odot }$ for very rich systems. The cluster
morphology is usually dominated by a regular centrally peaked main component
\cite{ReBo02,Ar05}.

As clusters are "dark matter" dominated objects, their formation and
evolution is driven by gravity. The mass function of the clusters is
determined by the initial conditions of the mass distribution set in the
early universe. The evolution of the large scale matter distribution on
scales comparable to the size of the clusters is linear. The overall process
of the gravitational growth of the density fluctuations and the development
of gravitational instabilities leading to cluster formation has been
extensively studied by using both analytical and numerical methods \cite
{Sch01}.

An other important component is the intra-cluster gas, which is
assumed to be isothermal, and in hydrostatic equilibrium. Most of
the baryonic mass in the cluster is contained in the gas. The
properties of the gas are important
observational quantities, since the total gravitational mass of the cluster $%
M_{tot}$ is determined by using the radial gas distribution \cite{ReBo02}.
By assuming that the density and pressure of the gas are $\rho _g$ and $p_g$%
, respectively, and by neglecting the quadratic terms in the field
equations, the basic equations describing the structure of a galactic
cluster in the brane world model are given by
\begin{equation}
-e^{-\lambda }\left( \frac{1}{r^{2}}-\frac{\lambda ^{\prime }}{r}\right) +%
\frac{1}{r^{2}}=8\pi G\rho _{g}+3\alpha U+\Lambda ,  \label{f11}
\end{equation}
\begin{equation}
e^{-\lambda }\left( \frac{\nu ^{\prime }}{r}+\frac{1}{r^{2}}\right) -\frac{1%
}{r^{2}}=8\pi Gp_{g}+\alpha \left( U+2P\right) -\Lambda ,  \label{f22}
\end{equation}
\begin{equation}
e^{-\lambda }\frac{1}{2}\left( \nu ^{\prime \prime }+\frac{\nu ^{\prime 2}}{2%
}+\frac{\nu ^{\prime }-\lambda ^{\prime }}{r}-\frac{\nu ^{\prime }\lambda
^{\prime }}{2}\right) =8\pi Gp_{g}+\alpha \left( U-P\right) -\Lambda ,
\label{f33}
\end{equation}
\begin{equation}
\nu ^{\prime }=-\frac{U^{\prime }+2P^{\prime }}{2U+P}-\frac{6P}{r\left(
2U+P\right) }-2\left( 4\pi G\right) \frac{\rho _{g}+p_{g}}{2U+P}\rho
_{g}^{^{\prime }}.  \label{f44}
\end{equation}

Eq.~(\ref{f11}) can immediately be integrated to give
\begin{equation}
e^{-\lambda }=1-\frac{C}{r}-\frac{2GM_{g}}{r}-\frac{GM_{U}\left( r\right) }{r%
}-\frac{\Lambda }{3}r^{2},  \label{m1}
\end{equation}
where $C$ is an arbitrary constant of integration, $M_{U}$ is defined
according to Eq.~(\ref{darkmass}) and $M_{g}$ is the total mass of the gas.

By substituting $\nu ^{\prime }$ given by Eq.~(\ref{f44}) into Eq.~(\ref{f22}%
) and with the use of Eq.~(\ref{m1}) we obtain the following system of
differential equations satisfied by the gas density, pressure and the dark
radiation term $U$, the dark pressure $P$ and the dark mass $M_{U}$,
respectively, describing the gravitational field of a cluster of galaxies in
the brane world model:
\begin{equation}
\frac{dM_{g}}{dr}=4\pi \rho _{g}r^{2}.  \label{eg}
\end{equation}
\begin{equation}
\frac{dM_{U}}{dr}=\frac{3\alpha }{G}r^{2}U.  \label{e2}
\end{equation}
\begin{eqnarray}
&&\frac{dU}{dr} =-\frac{\left( 2U+P\right) \left\{ C+2GM_{g}+GM_{U}+\left[
\alpha \left( U+2P\right) +8\pi p_{g}\right] r^{3}\right\} -\frac{2}{3}%
\Lambda r^{3}}{r^{2}\left( 1-\frac{C}{r}-\frac{2GM_{g}}{r}-\frac{%
GM_{U}\left( r\right) }{r}-\frac{\Lambda }{3}r^{2}\right) }-  \nonumber \\
&&2\frac{dP}{dr}-\frac{6P}{r}+ 2\left( 4\pi G\right) \frac{\rho _{g}+p_{g}}{%
2U+P}\rho _{g}^{^{\prime }}.
\end{eqnarray}

Since the most important contribution to the structure and dynamics of the
clusters of galaxies comes from the ''dark matter'', which we interpret as a
multi-dimensional effect, described by the contribution of the Weyl tensor
from the bulk, in discussing the properties of the clusters we can neglect,
in the first approximation, the effect of the ordinary matter, by taking $%
\rho _{g}\approx 0$ and $p_{g}\approx 0$, respectively.

Therefore the metric components $\nu (r)$ and $\lambda (r)$ inside the
spherically symmetric static cluster are related to the dark radiation $U$
and to the dark pressure $P$ via the Einstein gravitational field equations
and the effective energy-momentum tensor conservation equation, which in the
vacuum take the form~\cite{Ha03,Ma04}
\begin{equation}
\frac{dM_{U}}{dr}=\frac{3\alpha }{G}r^{2}U.  \label{ne1}
\end{equation}
\begin{equation}
\frac{dU}{dr}=-\frac{\left( 2U+P\right) \left\{ C+GM_{U}+\left[
\alpha
\left( U+2P\right) \right] r^{3}\right\} -\frac{2}{3}\Lambda r^{3}}{%
r^{2}\left( 1-\frac{C}{r}-\frac{GM_{U}\left( r\right) }{r}-\frac{\Lambda }{3}%
r^{2}\right) }-2\frac{dP}{dr}-\frac{6P}{r}.  \label{ne2}
\end{equation}

The system of equations (\ref{ne1}) and (\ref{ne2}) can be transformed to an
autonomous system of differential equations by means of the transformations
\begin{equation}
q=\frac{C}{r}+\frac{GM_{U}}{r}+\frac{\Lambda }{3}r^{2},\quad \mu =3\alpha
r^{2}U+3r^{2}\Lambda ,\quad p=3\alpha r^{2}P-3r^{2}\Lambda ,\quad \theta
=\ln r.  \label{trans}
\end{equation}

We shall call $\mu $ and $p$ the ``reduced'' dark radiation and pressure,
respectively.

With the use of the new variables given by Eqs.~(\ref{trans}), the
system of Eqs.~(\ref {ne1}) and (\ref{ne2}) become
\begin{equation}
\frac{dq}{d\theta }=\mu -q,  \label{aut1}
\end{equation}
\begin{equation}
\frac{d\mu }{d\theta }=-\frac{\left( 2\mu +p\right) \left[ q+\frac{1}{3}%
\left( \mu +2p\right) \right] }{1-q}-2\frac{dp}{d\theta }+2\mu -2p,
\label{aut2}
\end{equation}

In order to close the system of equations Eqs.~(\ref{aut1}) and (\ref{aut2}%
), we need to specify the ''equation of state'' of the dark pressure. As a
possible equation of state relating $p$ and $\mu $ we will assume a general
linear relation of the form
\begin{equation}
p(\mu )=\left( \Gamma -2\right) \mu +B,
\end{equation}
with $\Gamma $ and $B$ arbitrary constants. Therefore Eq.~(\ref{aut2}) takes
the form
\begin{equation}
\frac{d}{d\theta }\left[ \left( 2\Gamma -3\right) \mu \right] =-\frac{\left(
\Gamma \mu +B\right) \left[ q+\left( 2\Gamma -3\right) \mu /3+2B/3\right] }{%
1-q}+2\left( 3-\Gamma \right) \mu -2B,
\end{equation}
where we have neglected the possible effect of the cosmological constant on
the structure of the cluster.

For clusters of galaxies with virial masses of the order of $M_{V}\approx
10^{14}M_{\odot }$ and virial radii of the order of $R_{V}\approx 2$ Mpc,
the quantity $GM/R$ is of the order of $2\times 10^{-6}$. Since observations
show that inside the cluster the mass is a linearly increasing function of
the radius $r$, the value of this ratio is roughly the same at all points in
the cluster. Therefore from its definition it follows that generally $q<<1$,
and $1-q\approx 1$. Moreover, the quantities $q^{2}$ and $qdq/d\theta $ are
also very small as compared to $q$. Eq. (\ref{aut1}) gives $\mu
=q+dq/d\theta $, $d\mu /d\theta =dq/d\theta +d^{2}q/d\theta ^{2}$. Hence, by
neglecting the second order terms, we obtain for $q$ the following
differential equation:
\begin{equation}
\frac{d^{2}q}{d\theta ^{2}}+m\frac{dq}{d\theta }+nq=b,  \label{lineq}
\end{equation}
where we denoted
\begin{equation}
m=1+\frac{B}{3}+\frac{2\Gamma \left( B+3\right) -18}{3\left( 2\Gamma
-3\right) },
\end{equation}
\begin{equation}
n=\frac{B}{3}+\frac{2\left( B+3\right) \Gamma +3\left( B-6\right) }{3\left(
2\Gamma -3\right) },
\end{equation}
and
\begin{equation}
b=\frac{2B(B+3)}{3\left( 3-2\Gamma \right) },
\end{equation}
respectively. The general solution of Eq. (\ref{lineq}) is given by
\begin{equation}
q\left( \theta \right) =q_{0}+C_{1}e^{l_{1}\theta }+C_{2}e^{l_{2}\theta },
\end{equation}
where $C_{1}$ and $C_{2}$ are arbitrary constants of integration, and we
denoted
\begin{equation}
q_{0}=\frac{b}{n},
\end{equation}
and
\begin{equation}
l_{1,2}=\frac{-m\pm \sqrt{m^{2}-4n^{2}}}{2},  \label{l}
\end{equation}
respectively. The reduced dark radiation term is given by
\begin{equation}
\mu \left( \theta \right) =q_{0}+C_{1}\left( 1+l_{1}\right) e^{l_{1}\theta
}+C_{2}\left( 1+l_{2}\right) e^{l_{2}\theta }.
\end{equation}

For a positive $m$ from Eq.~(\ref{l}) it follows that both $l_{1}$ and $%
l_{2} $ are negative numbers, $l_{1}<0$ and $l_{2}<0$, respectively. In the
original radial variable $r$ we obtain for the dark radiation and the mass
distribution inside the cluster the expressions\bigskip
\begin{equation}
3\alpha U(r)=\frac{q_{0}}{r^{2}}+C_{1}\left( 1+l_{1}\right)
r^{l_{1}-2}+C_{2}\left( 1+l_{2}\right) r^{l_{2}-2},  \label{UDARK}
\end{equation}
and
\begin{equation}
GM_{U}(r)=q_{0}r\left( 1+\frac{C_{1}}{q_{0}}r^{l_{1}}+\frac{C_{2}}{q_{0}}%
r^{l_{2}}-\frac{C}{q_{0}r}\right) ,  \label{UMASS}
\end{equation}
respectively, where we have neglected the possible effect of the
cosmological constant.

The metric coefficient $\nu $ can be calculated from the equation
\begin{equation}
\nu ^{\prime }\left( \theta \right) =q\left( \theta \right) +\frac{2\Gamma -3%
}{3}\mu \left( \theta \right) +\frac{2B}{3},
\end{equation}
giving
\begin{equation}
e^{\nu \left( r\right) }=C_{3}r^{2\left( q_{0}\Gamma +B\right) /3}\exp \left[
C_{1}\frac{3+\left( 2\Gamma -3\right) \left( 1+l_{1}\right) }{3l_{1}}%
r^{l_{1}}+C_{2}\frac{3+\left( 2\Gamma -3\right) \left( 1+l_{2}\right) }{%
3l_{2}}r^{l_{2}}\right] ,  \label{NU}
\end{equation}
where $C_{3}$ is an arbitrary integration constant. In the limit of large
distances $e^{\nu \left( r\right) }$ behaves like $e^{\nu \left( r\right)
}\approx C_{3}r^{2\left( q_{0}\Gamma +B\right) /3}$. Inside the cluster we
can approximate $e^{-\lambda }\approx 1-C/r-GM_{U}(r)/r$.

The first term in the mass profile of the dark mass given by Eq.~(\ref{UMASS}%
) is linearly increasing with $r$, thus having a similar behavior to the
dark matter in clusters of galaxies. In order to obtain a model consistent
with observations, the second and third terms in Eq.~(\ref{UMASS}) must be
decreasing with increasing $r$. Therefore the constants $l_{1}$ and $l_{2}$
must also satisfy the conditions $l_{1}+1<0$ and $l_{2}+1<0$, respectively.
By assuming that $m$ and $n$ are positive numbers, the conditions $l_{1}<0$
and $l_{2}<0$ are realized automatically.

In the particular case $B=0$, we have $q_{0}\equiv 0$, and the solution is
defined only for values of $\Gamma $ so that $\Gamma \geq 21/8$. This shows
that a radiation like equation of state of the form $P=U/3$ is not allowed
in the present model. Moreover, the obtained mass profile is not consistent
with the observations, and for such an equation of state the dark radiation
cannot play the role of the dark matter. For the case $\Gamma =4$,
corresponding to an equation of state $P=2U$, the dark mass is given by $%
GM_{U}(r)=C_{1}/r^{1.97}+C_{2}/r^{0.82}-C$.

Since according to our physical interpretation $M_{U}$ is an
effective, geometry induced mass, it must satisfy the condition
$M_{U}\geq 0$ for all $r$. Therefore the solution obtained in the
present Section is valid only for values of the coordinate radius
$r$ so that $q_{0}r+C_{1}r^{l_{1}+1}+C_{2}r^{l_{2}+1}-C\geq 0$,
where we
assume that all integration constants are non-zero. In the limit of small $r$%
, taking into account that $l_{1}+1<0$ and $l_{2}+1<0$, and by
assuming that the constant $C$ is small and can be neglected near
the origin of the cluster, we obtain
$C_{1}r^{l_{1}}+C_{2}r^{l_{2}}\approx 0 $. If both $C_{1}$ and
$C_{2}$ are strictly positive, then the dark mass diverges at the
center of the cluster, $\lim_{r\rightarrow 0}M_{U}(r)=\infty $. On
the other hand, if $C_{1}$ and $C_{2}$ have different signs, there
is a single point $r_{0}$, given by $r_{0}\approx\left(
C_{2}/C_{1}\right) ^{1/\left( l_{1}-l_{2}\right) }$, so that
$M_{U}(r_{0})\approx0$ and $M_{U}(r_{0})<0$, for $%
r<r_{0}$. In this case our solution has physically acceptable
properties only in the region $r\geq r_0$.

The solution of the brane world model Einstein field equations, given by
Eqs.~(\ref{UMASS}), (\ref{UDARK}) and {\ref{NU}) depends on five, equation
of state related or integration constants, $q_{0}$, $C$ and $C_{i}$, $%
i=1,2,3 $, respectively. These constants must be determined by using some
appropriate boundary conditions, or by using some other observable
quantities. }Without any loss in generality we can take the integration
constant $C_{3}=1$, since the metric tensor component $\exp {(\nu )}$ can be
arbitrarily re-parameterized by a transformation of the time coordinate.

An important observational quantity is the radial velocity dispersion $%
\sigma _{r}^{2}$, which is related to the total mass in the cluster by the
relation $GM_{V}=\sigma _{r}^{2}R_{V}.$ \cite{Ca97}. By assuming that $%
R_{U}\approx R_{V}$ gives, with the use of the generalized virial theorem, $%
M_{V}\left( R_{V}\right) \approx M_{U}\left( R_{V}\right) $, respectively.
For $r=R_{V}$ Eq. (\ref{UMASS}) becomes $M_{U}\left( R_{V}\right) \approx
q_{0}R_{V}$, thus giving
\begin{equation}
q_{0}\approx \sigma _{r}^{2}.  \label{eqs1}
\end{equation}

Observationally, the mass profiles of the clusters of galaxies are obtained
from the equation
\begin{equation}
GM_{tot}\left( r\right) =-\sigma _{r}^{2}\left( r\right) r%
\left[ \frac{d\ln \sigma _{r}^{2}(r)}{d\ln r}+\frac{d\ln n(r)}{d\ln r}+2\beta _{\sigma }%
\right],
\end{equation}
where $n(r)$ is the spatial galaxy number density and $\beta
_{\sigma }$ is the velocity anisotropy parameter \cite{Ca97}. This
equation has a very similar mathematical structure with Eq.
(\ref{UMASS}), with the total observed mass of the cluster
interpreted as the dark mass. Therefore the other constants in the
model can be, at least in principle, obtained by fitting the dark
radiation $U(r)$ and dark mass $M_{U}(r)$ with the observed
density and mass profiles of the dark matter in clusters of
galaxies.

\section{Astrophysical applications}

Astrophysical observations together with cosmological simulations have shown
that the virialized part of the cluster corresponds roughly to a fixed
density contrast $\delta \sim 200$ as compared to the critical density of
the universe, $\rho _{c}\left( z\right) $, at the considered redshift, so
that $\rho _V=3M_{V}/4\pi R_{V}^{3}=\delta \rho _{c}\left( z\right)$, where $%
\rho _V$ is the virial density, $M_{V}$ and $R_{V}$ are the virial
mass and radius, $\rho _{c}\left( z\right)
=h^{2}(z)3H_{0}^{2}/8\pi G$, and $h(z)$ is the Hubble parameter
normalized to its local value: $h^{2}(z)=\Omega _{m}\left(
1+z\right) ^{3}+\Omega _{\Lambda }$, where $\Omega _{m}$ is the
mass density parameter and $\Omega _{\Lambda }$ is the dark energy
density parameter \cite{Ar05}.

Once the integrated mass as a function of radius is determined for galaxy
clusters, a physically meaningful fiducial radius for the mass measurement
has to be defined. The radii commonly used are either $r_{200}$ or $r_{500}$%
. These radii are the radii within the mean gravitational mass density of
the matter $\left\langle \rho _{tot}\right\rangle =200\rho _{c}$ or $500\rho
_{c}$. A pragmatic approach to the virial mass is to use $r_{200}$ as the
outer boundary \cite{ReBo02}.

The numerical values of the radius $r_{200}$ are in the range $r_{200}=0.85$
Mpc (for the cluster NGC 4636) and $r_{200}=4.49$ Mpc (for the cluster
A2163). A typical value for $r_{200}$ is $2$ Mpc. The masses corresponding
to $r_{200}$ and $r_{500}$ are denoted by $M_{200}$ and $M_{500}$,
respectively. Usually it is assumed that $M_{V}=M_{200}$ and $R_{V}=r_{200}$
\cite{ReBo02}.

\subsection{Dark radiation, dark mass and dark radius from galactic cluster
observations}

In clusters of galaxies most of the baryonic mass is in the form of the
intra-cluster gas. The gas mass density $\rho _{g}$ distribution can be
fitted with the observational data by using the following expression for the
radial baryonic mass (gas) distribution \cite{ReBo02}
\begin{equation}
\rho _{g}(r)=\rho _{0}\left( 1+\frac{r^{2}}{r_{c}^{2}}\right) ^{-3\beta /2},
\label{dens}
\end{equation}
where $r_{c}$ is the core radius, and $\rho _{0}$ and $\beta $ are
(cluster-dependent) constants.

A static spherical system in collisionless equilibrium can be described by
the Jeans equation, which is given by \cite{Bi87}
\begin{equation}
\frac{d}{dr}\left[ \rho _{g}\sigma _{r}^{2}\right] +\frac{2\rho _{g}\left(
r\right) }{r}\left( \sigma _{r}^{2}-\sigma _{\theta ,\phi }^{2}\right)
=-\rho _{g}\left( r\right) \frac{d\Phi }{dr},
\end{equation}
where $\Phi \left( r\right) $ is the gravitational potential, $\sigma _{r}$
and $\sigma _{\theta ,\phi }$ are the mass-weighted velocity dispersions in
the radial and tangential directions, respectively. We assume that the gas
is distributed isotropically inside the clusters and therefore we take $%
\sigma _{r}=\sigma _{\theta ,\phi }$. The pressure profile $P_{g}$ is
related to the velocity dispersion by $P_{g}=\rho _{g}\sigma _{r}^{2}$.

Therefore the equation describing the isotropic equilibrium of the gas is
\begin{equation}
\frac{dP_{g}\left( r\right) }{dr}=-\rho _{g}\left( r\right) \frac{d\Phi
\left( r\right) }{dr}=-\frac{GM_{tot}\left( r\right) }{r^{2}}\rho _{g}\left(
r\right) ,  \label{equil}
\end{equation}
where $M_{tot}(r)$ is the total mass inside radius $r$. In obtaining Eq. (%
\ref{equil}) we have assumed that the gravitational field is weak, and that
the gravitational potential satisfies the usual Poisson equation $\Delta
\Phi \approx 4\pi \rho _{tot}$, with the total energy density $\rho
_{tot}\approx \rho _{g}+\rho _{m}+3\alpha U/8\pi G$ also including the
energy density $\rho _{m}$ of other forms of matter, different from gas,
like, for example, luminous matter, massive neutrinos etc., and the bulk
(five-dimensional) effects. In the following we will systematically neglect
the quadratic contribution to the energy-momentum tensor. The energy density
of the gas is much larger than the pressure, $\rho _{g}>>P_{g}$.

The observed X-ray emission from the hot, ionized intra-cluster gas is
usually interpreted by assuming that the gas is in isothermal equilibrium.
Therefore, we may assume that the pressure $P_{g}$ of the gas satisfies the
equation of state $P_{g}\left( r\right) =\left( k_{B}T_{g}/\mu m_{p}\right)
\rho _{g}\left( r\right) $, where $k_{B}$ is Boltzmann's constant, $T_{g}$
is the gas temperature, $\mu \approx 0.61$ \cite{ReBo02} is the mean atomic
weight of the particles in the cluster gas, and $m_{p}$ is the proton mass.
Then Eq. (\ref{equil}) gives
\begin{equation}
M_{tot}(r)=-\frac{k_{B}T_{g}}{\mu m_{p}G}r^{2}\frac{d}{dr}\ln \rho _{g}.
\end{equation}

By using the density profile of the gas given by Eq. (\ref{dens}) we obtain
for the mass profile inside the cluster the relation \cite{ReBo02}
\begin{equation}
M_{tot}(r)=\frac{3k_{B}\beta T_{g}}{\mu m_{p}G}\frac{r^{3}}{r_{c}^{2}+r^{2}}.
\label{mp}
\end{equation}

On the other hand, from the Einstein field equation Eq. (\ref{f1}) it
follows that the total mass inside the radius $r$ satisfies the following
mass continuity equation,
\begin{equation}
\frac{dM_{tot}\left( r\right) }{dr}=4\pi r^{2}\rho _{g}\left( r\right)
+3\alpha r^{2}U\left( r\right) ,  \label{massf}
\end{equation}
where we have neglected again the quadratic contributions in the gas density
and pressure to the total mass. Since the gas density and the total mass
profile inside the cluster are given by Eqs. (\ref{dens}) and (\ref{mp}),
respectively, we can obtain immediately the expression of the dark radiation
term inside the cluster as
\begin{equation}
3\alpha GU(r)=\left[ \frac{3k_{B}\beta T_{g}r^{2}\left(
r^{2}+3r_{c}^{2}\right) }{\mu m_{p}\left( r_{c}^{2}+r^{2}\right) ^{2}}-\frac{%
4\pi G\rho _{0}r^{2}}{\left( 1+r^{2}/r_{c}^{2}\right) ^{3\beta /2}}\right]
\frac{1}{r^{2}}.
\end{equation}

In the limit $r>>r_{c}$ we obtain for the dark radiation the simple relation
\begin{equation}
3\alpha GU(r)=\left[ \frac{3k_{B}\beta T_{g}}{\mu m_{p}}-4\pi G\rho
_{0}r_{c}^{3\beta }r^{2-3\beta }\right] \frac{1}{r^{2}}.
\end{equation}

The dark mass can be obtained generally as
\begin{equation}
GM_{U}(r)=3\alpha \int_{0}^{r}r^{2}U(r)dr=\left[ \frac{3k_{B}\beta T_{g}}{%
\mu m_{p}}\frac{r}{1+r_{c}^{2}/r^{2}}-4\pi G\rho _{0}\int_{0}^{r}\frac{%
r^{2}dr}{\left( 1+r^{2}/r_{c}^{2}\right) ^{3\beta /2}}\right] .
\end{equation}

In the limit $r>>r_{c}$ the dark mass $M_{U}$ is given as a function of $r$
by the expression
\begin{equation}
GM_{U}\left( r\right) \approx \left[ \frac{3k_{B}\beta T_{g}}{\mu m_{p}}-%
\frac{4\pi G\rho _{0}r_{c}^{3\beta }r^{2-3\beta }}{3\left( 1-\beta \right) }%
\right] r.  \label{GM}
\end{equation}

Let's assume first that we can neglect the contribution of the gas to the
dark radiation and dark mass, respectively. Then we obtain
\begin{equation}
3\alpha GU(r)\approx \left( \frac{3k_{B}\beta T_{g}}{\mu
m_{p}}\right) r^{-2},
\end{equation}
and
\begin{equation}
GM_{U}\left( r\right) \approx \left( \frac{3k_{B}\beta T_{g}}{\mu
m_{p}}\right) r,
\end{equation}
respectively.

There is a simple way to estimate an upper bound for the cutoff of the dark
mass. The idea is to consider the point at which the decaying density
profile of the dark radiation associated to the galaxy cluster becomes
smaller than the average energy density of the Universe. Let the value of
the coordinate radius at the point where the two densities are equal to be $%
R_{U}^{(cr)}$. Then at this point $3\alpha GU\left( R_{U}^{(cr)}\right)
=\rho _{univ}$, where $\rho _{univ}$ is the mean energy density of the
universe. By assuming $\rho _{univ}=\rho _{c}=3H^{2}/8\pi G=4.6975\times
10^{-30}h_{50}^{2}$ g/cm$^{-3}$, where $H=50h_{50}\mathrm{km}/\mathrm{Mpc}/%
\mathrm{s}$ \cite{ReBo02}, we obtain
\begin{equation}\label{Rucr}
R_{U}^{(cr)}=\left( \frac{3k_{B}\beta T_{g}}{\mu m_{p}G\rho _{c}}\right)
^{1/2}=91. 33\sqrt{\beta }\left( \frac{k_{B}T_{g}}{5\text{keV}}%
\right) ^{1/2}h_{50}^{-1}\mathrm{Mpc.}
\end{equation}

The total dark mass corresponding to this value is
\begin{equation}
M_{U}^{(cr)}=M_{U}\left( R_{U}^{(cr)}\right) =4.83\times
10^{16}\beta ^{3/2}\left( \frac{k_{B}T_{g}}{5\text{keV}}\right)
^{3/2}h_{50}^{-1}M_{\odot }.
\end{equation}

This value of the mass is consistent with the observations of the
mass distribution in the clusters of galaxies. However, according
to the brane world scenario, we predict that the dark mass and its
effects extends beyond the virial radius of the clusters, which is
of the order of only a few Mpc.

Let's now briefly consider the effect of the gas on the dark mass
distribution. If $\beta <2/3$, the dark mass has a maximum value, which
occurs at a radius $R_{1}$, which can be obtained from the condition $%
dM_{U}(r)/dr=0$, and is given by
\begin{equation}
R_{1}=\left( \frac{3k_{B}\beta T_{g}}{4\pi G\rho _{0}\mu m_{p}r_{c}^{3\beta }%
}\right) ^{1/\left( 2-3\beta \right) }.
\end{equation}

The maximum value of the dark mass can be found as
\begin{equation}
GM_{U}^{\max }\left( R_{1}\right) =\frac{\left( 2-3\beta \right)
}{3\left(
1-\beta \right) }\left( \frac{3k_{B}\beta T_{g}}{\mu m_{p}}\right) ^{\frac{%
3\left( 1-\beta \right) }{2-3\beta }}\left( 4\pi G\rho _{0}r_{c}^{3\beta
}\right) ^{-1/\left( 2-3\beta \right) }.
\end{equation}

The central density of the gas $\rho _{0}$ can be estimated from
the total gas mass $M_{g}$ as
\begin{equation}
\rho _{0}\approx \frac{3\left( 1-\beta \right) M_g\left(
R_{V}\right) }{4\pi r_{c}^{3}\left( R_{V}/r_{c}\right) ^{3\left(
1-\beta \right) }}.
\end{equation}
Hence we obtain
\begin{equation}\label{R1max}
R_{1}=\left[ \frac{k_{B}\beta T_{g}}{\left( 1-\beta \right)
GM_{g}\left( R_{V}\right) \mu m_{p}}\right] ^{1/\left( 2-3\beta
\right) }R_{V}^{3\left( 1-\beta \right) /\left( 2-3\beta \right)
},
\end{equation}
and
\begin{equation}
GM_{U}^{\max }\left( R_{1}\right) =\frac{ 2-3\beta }{ \left[
3\left(
1-\beta \right) \right]^{3\left(1-\beta \right) }} \left( \frac{%
3k_{B}\beta T_{g}}{\mu m_{p}}\right) ^{\frac{3\left( 1-\beta \right) }{%
2-3\beta }}\left[ GM_{g}\left( R_{V}\right) \right] ^{-1/\left(
2-3\beta \right) }R_{V}^{3\left( 1-\beta \right) /\left( 2-3\beta
\right) }.
\end{equation}

Usually the mass of the gas represent $5-10\%$ of the total mass
in the cluster \cite{Ca97,ReBo02}. For $\beta =1/2$,
$k_{B}T_{g}=5$ keV, and assuming for the mass of
the gas a value of the order $M_g=7\times 10^{13}M_{\odot}$, we obtain $%
R_{1}\approx 50.92$ Mpc and $M_{U}^{\max }\left( R_{1}\right)
=6.6\times 10^{16}M_{\odot}$, respectively. For $\beta \geq 2/3$
the mass is a linearly increasing function of the radius.

By comparing Eqs.~(\ref{Rucr}) and (\ref{R1max}) we predict that,
due to the presence of the gas in the cluster, the dark mass,
playing the role of the "dark matter", should cut-off earlier, and
before it merges into the background.

Astrophysical observations are extended and interpreted in terms
of an outer density of the order of $\delta \rho _{univ}$, with a
fixed density contrast $\delta =200$. In order to compare the
predictions of the brane world models with the observations it is
necessary to estimate the dark radius for an outer density of the
dark radiation of the order of $\delta \rho _{cr}$, with the
density contrast $\delta =200$. Hence, the dark radius
$R_{U}^{200}$ corresponding to the density contrast $\delta =200$
is given as a solution of the non-linear algebraic equation
\begin{equation}
\frac{3k_{B}\beta T_{g}}{\mu m_{p}}-4\pi G\rho _{0}\left( \frac{r_{c}}{%
R_{U}^{200}}\right) ^{3\beta }\left( R_{U}^{200}\right) ^{2}=\delta G\rho
_{c}\left( R_{U}^{200}\right) ^{2}.
\end{equation}

By evaluating the gas density at $R_{U}^{200}$ we obtain
\begin{equation}
\rho _{g}\left( R_{U}^{200}\right) \approx \rho _{0}\left( \frac{r_{c}}{%
R_{U}^{200}}\right) ^{3\beta }=\delta _{gd}\rho _{c},
\end{equation}
where $\delta _{gd}$ gives the density of the amount of the total mass in
form of the gas at a distance $R_{U}^{200}$ from the center of the cluster
with respect to the critical density. Then we obtain for the dark radius of
the cluster the expression
\begin{equation}
R_{U}^{200}=\sqrt{\frac{3k_{B}\beta T_{g}}{\mu Gm_{p}\rho _{c}\left( 4\pi
\delta _{gd}+\delta \right) }}=\frac{R_{U}^{(cr)}}{\sqrt{4\pi \delta
_{gd}+\delta }}=\frac{91. 33}{\sqrt{4\pi \delta _{gd}+\delta }}%
\sqrt{\beta }\left( \frac{k_{B}T_{g}}{5\text{keV}}\right) ^{1/2}h_{50}^{-1}%
\mathrm{Mpc}.
\end{equation}

For $k_{B}T_{g}=5$ keV, $\beta =2/3$ and $\delta _{gd}=20$ we obtain $%
R_{U}^{200}\approx 3.51$ Mpc. This shows that $R_{U}^{200}$ is of
the same order of magnitude as the virial radius $R_{V}=r_{200}$,
and the presence of the gas considerably decreases the magnitude
of the dark radius. Therefore it follows from the virial theorem
that the virial mass of the galaxy clusters is mainly determined
by the dark mass associated to the transmitted projection of the
bulk Weyl tensor.

\subsection{Radial velocity dispersion in galactic clusters}

The virial mass can also be expressed in terms of the characteristic
velocity dispersion $\sigma _{1}$ as \cite{Ca97}
\begin{equation}
M_{V}=\frac{3}{G}\sigma _{1}^{2}R_{V}.
\end{equation}

By assuming that the velocity distribution in the cluster is isotropic, we
have $\left\langle u^2\right\rangle=\left\langle
u_r^2\right\rangle+\left\langle u_{\theta }^2\right\rangle+\left\langle
u_{\phi }^2\right\rangle=3\left\langle u_r^2\right\rangle=3\sigma _r^2$,
with $\sigma _r^2$ the radial velocity dispersion. $\sigma _1$ and $\sigma
_r $ are related by $3\sigma _1^2=\sigma _r^2$.

In order to derive the radial velocity dispersion relation for clusters of
galaxies in brane world models we start from Eq. (\ref{tetr2}). Taking into
account that the velocity distribution is isotropic, we obtain
\begin{equation}
\frac{d}{dr}\left( \rho \sigma _{r}^{2}\right) +\frac{1}{2}\rho \frac{d\nu }{%
dr}=0.  \label{veldis}
\end{equation}

Since inside the cluster $e^{-\lambda }\approx 1$, by neglecting the
cosmological constant the Einstein field equation Eq. (\ref{ff}) becomes
\begin{equation}
\frac{\nu ^{\prime \prime }}{2}+\frac{\nu ^{\prime }}{r}=3\alpha U+4\pi
G\rho ,
\end{equation}
and can be written in the equivalent form
\begin{equation}
\frac{1}{2r^{2}}\frac{d}{dr}\left( r^{2}\nu ^{\prime }\right) =3\alpha
U+4\pi G\rho ,
\end{equation}
giving upon integration
\begin{equation}
r^{2}\nu ^{\prime }=2GM_{U}(r)+2GM(r)+2C_{4},
\end{equation}
where $C_{4}$ is an arbitrary constant of integration. Since from Eq. (\ref
{veldis}) we have $\nu ^{\prime }=-(2/\rho )d\left( \rho \sigma
_{r}^{2}\right) /dr$, it follows that the radial velocity dispersion of the
galactic clusters on the brane satisfies the differential equation
\begin{equation}
\frac{d}{dr}\left( \rho \sigma _{r}^{2}\right) =-\frac{GM_{U}(r)}{r^{2}}\rho
(r)-\frac{GM(r)}{r^{2}}\rho (r)-\frac{C_{4}}{r^{2}}\rho (r),
\end{equation}
with the general solution given by
\begin{equation}
\sigma _{r}^{2}(r)=-\frac{1}{\rho }\int \left[ \frac{GM_{U}(r)}{r^{2}}\rho
(r)+\frac{GM(r)}{r^{2}}\rho (r)+\frac{C_{4}}{r^{2}}\rho (r)\right] dr+\frac{%
C_{5}}{\rho },  \label{veldisf}
\end{equation}
where $C_{5}$ is an integration constant.

As an example of the application of Eq. (\ref{veldisf}) we consider the case
in which the density $\rho $ of the normal matter inside the cluster has a
power law distribution, so that
\begin{equation}
\rho (r)=\rho _{0}r^{-\gamma },
\end{equation}
with $\rho _{0}$ and $\gamma \neq 1,3$ positive constants. The corresponding
normal matter mass profile is $M(r)=4\pi \rho _{0}r^{3-\gamma }/\left(
3-\gamma \right) $. The dark mass $GM_{U}=q_{0}r$ is linearly proportional
to $r$, as has been shown in the previous Section. Therefore after some
simple calculations we obtain
\begin{equation}
\sigma _{r}^{2}(r)=\frac{q_{0}}{\gamma }+\frac{2\pi G\rho _{0}}{\left(
\gamma -1\right) \left( 3-\gamma \right) }r^{2-\gamma }+\frac{C_{4}}{\gamma
+1}\frac{1}{r}+\frac{C_{5}}{\rho _{0}}r^{\gamma },\gamma \neq 1,3.
\end{equation}

In the case $\gamma =1$ we obtain

\begin{equation}
\sigma _{r}^{2}(r)=q_{0}+\frac{C_{4}}{2r}-2\pi G\rho _{0}r\ln r+\frac{C_{5}}{%
\rho _{0}}r,\gamma =1.
\end{equation}

For $\gamma =3$ we find

\begin{equation}
\sigma _{r}^{2}(r)=\frac{q_{0}}{3}-\pi G\rho _{0}\left( \ln r+\frac{1}{4}%
\right) \frac{1}{r^{4}}+\frac{C_{4}}{4}\frac{1}{r}+\frac{C_{5}}{\rho _{0}}%
r^{3},\gamma =3.
\end{equation}

Usually the observed data for the velocity dispersion in clusters of
galaxies are analyzed by assuming the simple form $\sigma
_{r}^{2}(r)=B/(r+b) $ for the radial velocity dispersion, with $B$ and $b$
constants. As for the density of the galaxies in the clusters the relation $%
\rho \left( r\right) =A/r\left( r+a\right) ^{2}$, with $A$ and $a$
constants, is used. The data are then fitted with these functions by using a
non-linear fitting procedure \cite{Ca97}. For $r<<a$, $\rho (r)\approx A/r$,
while for $r>>a$, $\rho (r)$ behaves like $\rho (r)\approx A/r^{3}$.
Therefore the comparison of the observed velocity dispersion profiles of the
galaxy clusters and the velocity dispersion profiles predicted by the brane
world scenario may give a powerful method to discriminate between the
different theoretical models.

\section{Discussions and final remarks}

The galactic rotation curves and the mass distribution in clusters of
galaxies continue to pose a challenge to present day physics. One would like
to have a better understanding of some of the intriguing phenomena
associated with them, like their universality and the very good correlation
between the amount of dark matter and the luminous matter in the galaxy. To
explain these observations, the most commonly considered models are based on
particle physics in the framework of Newtonian gravity, or of some
extensions of general relativity \cite{dark}.

In the present paper we have considered, and further developed, an
alternative view to the dark matter problem
\cite{Ma04,Ha04,Ha05,BoHa07}, namely, the possibility that the
galactic rotation curves and the mass discrepancy in clusters of
galaxies can naturally be explained in models in which our
Universe is a domain wall (a brane) in a multi-dimensional
space-time (the bulk). The extra-terms in the gravitational field
equations on the brane induce a supplementary gravitational
interaction, which can account for the observed behavior of the
galactic rotation curves. By using the simple observational fact
of the constancy of the galactic rotation curves, the galactic
metric and the corresponding Weyl stresses (dark radiation and
dark pressure) can be completely reconstructed \cite{Ha05,BoHa07}.

In order to estimate the effect of the bulk effects on the
extra-galactic scale, at the level of the clusters of galaxies, we
have generalized the virial theorem to include the contribution of
the non-compact extra dimensions. To derive the generalized virial
theorem we have used a method based on the collisionless Boltzmann
equation. The generalized virial theorem allows to remove the
virial mass discrepancy, by showing that the virial mass $M_{V}$
is proportional, as one can see from Eq. (\ref{virial}), to the
dark mass $M_{U}$ associated to the dark radiation term $U$, which
is the scalar component of the projected Weyl tensor $E_{\mu \nu
}$. The dark mass can be directly related to the observed virial
mass, and, for typical clusters of galaxies, it is of the same
order of magnitude as the virial mass. This shows that at the
extra-galactic scale the dark mass plays the role of what is
conventionally called the dark matter.

In the present model the cut-off of the dark mass (which
represents the dark matter in the conventional interpretation) is
not due to the merging of the dark radiation with the cosmological
background, but it is due to the presence of the intra-cluster
gas, which is the only form of conventional matter in our theory.

In order to derive our main results we have used a set of
approximations and assumptions, which can be classified as a)
brane world model related b) small velocity limit approximation
and c) weak gravitational field approximation, respectively. In
our treatment of the brane world models we have systematically
ignored the second order terms in the matter density. This
approximation is justified since the ratio $\rho /\lambda _b$,
where $\lambda _b$ is the brane tension, is very small in the case
of the galaxy clusters. By taking for the mean density of the
baryonic matter in the cluster a value of $\rho \approx
10^{-20}-10^{-23}$ g/cm$^{3}$ \cite{ReBo02,Ar05,Sch01}, and for
the brane tension a value of the order of $\lambda _b=(100\; {\rm
GeV})^4=1.37\times 10^{21}$ g/cm$^{3}$ \cite{GeMa01}, it follows
that the ratio $\rho /\lambda _b$ is indeed negligible small at
the level of the clusters of galaxies, $\rho /\lambda _b<<1$, and
hence the second order effects in the matter density can be safely
neglected. Since the dispersion of the velocity of the galaxies in
the clusters is of the order $600-1000$ km/s
\cite{ReBo02,Ar05,Sch01}, giving for the square of the ratio of
the velocity and of the speed of light a value of the order of
$\left(v/c\right)^2\approx 4\times 10^{-6}$ - $1.11\times
10^{-5}<<1$, we can neglect the relativistic effects in the
Boltzmann equation, and use the small velocity limit of it. The
intensity of the gravitational effects can be estimated from the
ratio $GM/R$, which for typical clusters is of the order of
$10^{-6}<<1$. Therefore inside galactic clusters the gravitational
field is weak, and the linear approximation of the gravitational
equations provides an accurate description of the dynamics of the
massive test bodies. We have also systematically neglected the
possible effects of the cosmological constant. As for the physical
structure of the clusters, we have assumed that they are dark
radiation dominated spherically symmetric static astrophysical
systems, with a normal baryonic matter content representing only
$5-10\%$ of the total equivalent mass energy density. Both these
assumptions are very well supported by observations and
cosmological simulations \cite{ReBo02,Ar05,Sch01}.

In the present model all the relevant physical quantities,
including the dark mass and the dark radius $R_{U}$, which
describe the non-local effects due to the gravitational field of
the bulk, are expressed in terms of observable parameters (the
temperature and the virial radius of the cluster, and the observed
velocity dispersion). Hence it follows that the contributions of
the bulk Weyl tensor in the case of the clusters of galaxies on
the brane - the dark mass $M_{U}$, the dark radius $R_{U}$, and
the gravitational energy $\Omega _{U}$ - can be determined from
the study of the X-ray properties of the gas within the virial
radius.

Generally, the virial mass $M_V$ is obtained from the
observational study of the velocity dispersions of the stars in
the cluster. Thus, it cannot give a reliable estimation of the
numerical value of the total mass $M_g+M_U$ in the cluster. A very
useful method for the study of the total mass distribution in
clusters is the gravitational lensing of the light, which may
provide direct evidence for the total mass distribution, and for
the gravitational effects at large distances from the cluster.

Therefore, this opens the possibility of testing the brane world
models by using astronomical and astrophysical observations at the
extra-galactic scale. In this paper we have provided some basic
theoretical tools necessary for the in depth comparison of the
predictions of the brane world model and of the observational
results.

\section*{Acknowledgements}

This work is supported by the RGC grant No. 7027/06P of the government of
the Hong Kong SAR.

\end{document}